\documentclass[12pt]{iopart}
\usepackage{graphicx}
\usepackage{dcolumn}
\usepackage{bm}
\usepackage{cite}
\usepackage{amssymb}
\usepackage{epsfig}
\usepackage{multirow}
\usepackage{graphics}
\usepackage{color}

\begin{document}

\title[Importance of vdW interaction on thermodynamics properties of NaCl]{Importance of van der Waals interaction on structural, vibrational, and thermodynamics properties of NaCl}

\author{Michel L. Marcondes$^{1,2}$, Renata M. Wentzcovitch$^{1,3}$, Lucy  V. C. Assali$^2$}
\address{$^1$ Department of Earth and Environmental Sciences, Columbia University, Lamont-Doherty Earth Observatory, Palisades, NY 10964, USA}
\address{$^2$ Instituto de F\'{\i}sica, Universidade de S\~ao Paulo, CEP 05508-090, S\~ao Paulo, SP, Brazil}
\address{$^3$ Department of Applied Physics and Applied Mathematics, Columbia University, New York, NY 10027, USA}

\date{\today}

\begin{abstract}

Thermal equations of state (EoS) are essential in several scientific domains. However, experimental determination of EoS parameters may be limited at extreme conditions, therefore, {\it ab~initio} calculations have become an important method to obtain them. Density Functional Theory (DFT) and its extensions with various degrees of approximations for the exchange and correlation (XC) energy is the method of choice, but large errors in the EoS parameters are still common. The alkali halides have been problematic from the onset of this field and the quest for appropriate DFT functionals for such ionic and relatively weakly bonded systems has remained an active topic of research. Here we use DFT + van der Waals functionals to calculate vibrational properties, thermal EoS, thermodynamic properties, and the $B$1 to $B$2 phase boundary of NaCl. Our results reveal i) a remarkable improvement over the performance of standard Local Density Approximation and Generalized Gradient Approximation functionals for all these properties and phase transition boundary, as well as ii) great sensitivity  of anharmonic effects on the choice of XC functional.

\end{abstract}

\maketitle

\section{Introduction}
\label{sec1}

Reliable high pressure and high temperature (PT) equations of state (EoS) are fundamental for addressing materials properties at extreme conditions, from planetary sciences to engineering applications. Experiments can be challenging at high PT and \textit{ab~initio} calculations based on Density Functional Theory (DFT) \cite{Hohenberg1964, Kohn1965} have become an indispensable tool to investigate materials in such conditions. Although DFT is a predictive approach, the absence of an exact description of the exchange-correlation (XC) energy produces well-known systematic deviations from experimental data. Although the inclusion of vibrational effects greatly improves the description of strongly bonded ionic materials, such as silicates and oxides \cite{Karki1999,Wentzcovitch2010a}, systematic deviations can still be recognized in high temperature calculations. 

Rock-salt NaCl is an important and widely studied material \cite{Liu2010a,Xiong2014,Ono2006,Ono2008,Brown1999a}, whose high pressure thermoelastic properties has gained renewed attention, especially after the discovery of great pre-salt oil basins in several coastal regions \cite{Hudec2007,Marcondes2015b}. Accurate thermal EoSs are highly desirable for seismic mapping of these fields and for a precise extension of the NaCl pressure scale. Decker was the first to develop a NaCl pressure scale\cite{Decker1965,Decker1966,Decker1971a}, which was further improved by Brown \cite{Brown1999a}. Fritz \textit{et al.} measured the Hugoniot up to 25 GPa at 300 K \cite{Fritz1972a}, Boehler and Kennedy measured the thermal EoS up to 3.5 GPa and 770 K using a piston cylinder apparatus \cite{Boehler1980a}, and semiempirical EoSs were constructed by Dorogokupets \cite{Doro2007} and Matsui \cite{Matsui2012}. A hybrid \textit{ab~initio}/experimental correction scheme was also developed and produced accurate NaCl EoS parameters and thermoelastic properties\cite{Marcondes2015a,Marcondes2015b}. These recent efforts reaffirmed the problematic nature of DFT calculations for NaCl, which has been evident since it was first investigated \cite{Froyen1986}. In general, EoS for alkali halides are not well calculated when compared to strongly bonded ionic compounds with the same crystalline structure. For instance, MgO EoS parameters at 300 K deviate by at most 0.6\% from experimental values   \cite{Wu2008}, while for NaCl this deviation can be 6\% \cite{Marcondes2015a}.

The quest for appropriate DFT functionals to describe such ionic and relatively weakly bonded systems has remained an active topic of research. The popular Local Density Approximation (LDA) \cite{Kohn1965,Perdew1981a} and Generalized Gradient Approximation (GGA) \cite{Perdew1996a} functionals are unable to describe the van der Waals (vdW) interaction correctly and, given the nearly closed shell nature of the Na$^+$ and Cl$^-$ ions, it is natural to ask how important the vdW interaction really is. In recent years, several reviews have been written about DFT based methods that include the vdW interaction \cite{Grimme2007,Johnson2009,Klimes2012,Berland2015,Grimme2016,Hermann2017}. The DFT-D and DFT-D2 methods add an attractive dispersion term of the form $-C_6/r^6$ to the total energy. The DFT-D method first included the dispersion energy in the Hartree-Fock equations\cite{Hepburn1975, Ahlrichs1977}, but it was further extended later to DFT\cite{Wu2001, Elstner2001, Wu2002, Zimmerli2004, Grimme2004, Grimme2006b}. This method led to the development of the DFT-D3 method that takes into account the dependence of the $C_6$ parameter on the environment\cite{Grimme2006b,Grimme2010,Tkatchenko2009,Tkatchenko2012, Bucko2013a}. The vdW energy is a correlation effect; therefore it is included in the exact XC term in DFT. The Rutgers-Chalmers family of van der Waals density functionals (vdW-DF) reconstructs the correlation energy to incorporate the long and medium range interactions\cite{Rydberg2003,Dion2004, Langreth2005, Thonhauser2007, Vydrov2009, Vydrov2010, Lee2010,  Klimes2011, Berland2015}. Several investigations have addressed the importance of vdW energy to properly describe static properties of partially covalent/ionic solids\cite{Zhang2011,Klimes2011,Moellmann2012,Bucko2013b,Park2015}. Particularly for NaCl, the inclusion of vdW interaction energy resulted in more accurate lattice parameter and cohesive energy, compared to experiments\cite{Zhang2011,Klimes2011,Bucko2013b}. Nonetheless, these studies did not address the effect of finite temperatures on the EoS parameters. Consideration of vibrational/thermal effects is essential to address the true predictive power of a functional. Using the quasiharmonic approximation (QHA), here we extend investigations of vdW functionals' performance to compute vibrational and thermodynamics properties of the NaCl, including the $B1$ to $B2$ phase transformation boundary. We show that LDA and GGA functionals, augmented by the dispersive vdW interaction result in a remarkable improvement in the calculated thermal EoS parameters and in other vibrational and thermodynamic properties.

\section{Methods}
\label{sec2}

\textit{Ab~initio} calculations were performed using VASP \cite{vasp1}, Wien2k \cite{Wien2k}, and Quantum ESPRESSO \cite{Espresso,Espresso2} software.
The XC energy was calculated using the LDA, PBE (GGA), and the vdW-DF within the Dion \textit{et al.} scheme\cite{Dion2004}. In VASP calculations, the electronic wavefunctions were expanded using the Projected Augmented Wave (PAW) method \cite{Blochl1994, Kresse1999}. The plane wave cutoff used was 600 eV and the Brillouin zone sampling for electronic states was performed on a displaced 6$\times$6$\times$6 \textbf{k}-mesh.  
Thermodynamics properties were obtained using the QHA\cite{Born1956,Carrier2007}, where the Helmholtz free energy is given by:
\begin{eqnarray}
F\left(V,T\right) = E(V) + \frac{1}{2}\sum_{q,m}\hbar\omega_{q,m}(V) \nonumber \\
 + k_BT\sum_{q,m}\ln\left\{1 - \exp\left[-\frac{\hbar\omega_{q,m}(V)}{k_BT}\right]\right\}.
\label{helm}
\end{eqnarray}
The first term on the right-hand side is the DFT static energy; the second and third terms are the zero point motion and thermal energies, respectively. Vibrational frequencies were computed using the finite displacement method\cite{Srivastava1990,Kresse1995,Alfe2009} with the Phonopy code \cite{Togo2015} using a 4$\times$4$\times$4 supercell and interpolated in a 12$\times$12$\times$12 \textbf{q}-mesh to produce the vibrational density of states (VDoS)\cite{Wang2010}. The EoS parameters were obtained by fitting the third order finite strain Birch-Murnaghan EoS to the energy vs. volume relation. The Rutgers-Chalmers vdW-DF family of functionals has produced good lattice constants, bulk moduli and atomization energies of solids\cite{Lee2010, Klimes2011}, compared to experimental data. Therefore we evaluate the efficiency of this method to compute all thermodynamics properties of NaCl. In this scheme, the correlation energy, responsible for the vdW interaction is split into two terms:
\begin{equation}
E_c[n] = E_c^{LDA}[n] + E_c^{nl}[n],
\end{equation}
where $E_c^{LDA}$ is the LDA correlation energy and $E_c^{nl}$ is a correlation term that incorporates a long range and a non-local interaction. For the exchange energy, the original vdW-DF from Dion\cite{Dion2004} used the revPBE\cite{Zhang1998} functional, but other functionals have also been developed. In particular, the optimized functionals from Klimes \textit{et al.} produced good results for static properties of solids\cite{Klimes2010,Klimes2011}. Here, we extend these investigations by calculating the effect of vdW corrections in the vibrational and thermodynamic properties of NaCl, using the vdW-DF method with different functionals for the exchange energy. 
Although beyond the scope of this work, other functionals should also be tested, e.g., the strongly constrained and appropriately normed (SCAN) functional\cite{Sun2015} that obeys all known constraints of a meta-GGA that was combined with the rVV10 vdW functional\cite{Vydrov2010,Sabatini2013} showing promising results\cite{Peng2016}.

\section{Results and discussion}
\label{sec3}
\subsection{Structural, vibrational, and thermodynamic properties of $B1$-type NaCl}
\begin{figure*}
\begin{center}
\includegraphics[scale=0.35]{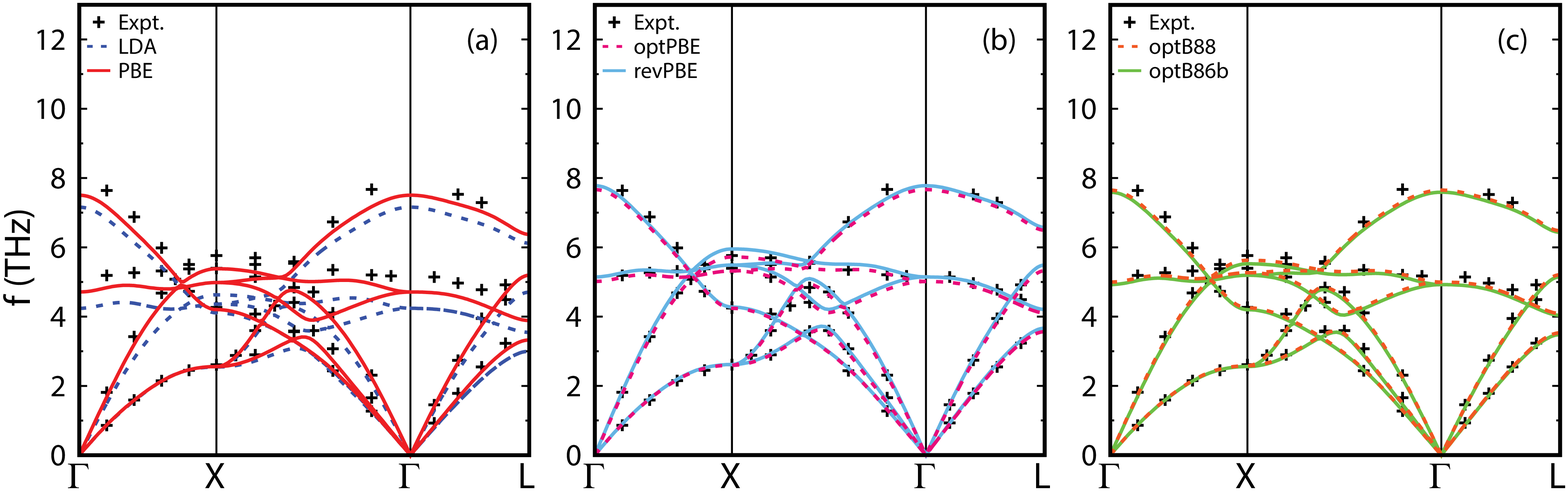}
\caption{Phonon dispersions of $B1$-type NaCl computed using (a) LDA and PBE functionals; (b) revPBE and optPBE vdW functionals; (c) optB88 and optB86b vdW functionals. Experimental data ($+$) are from  reference \cite{Raunio1969}. Calculations were performed at the experimental volume \cite{Boehler1980a}.}
\label{fig1}
\end{center}
\end{figure*}

The influence of vdW corrections in the NaCl thermal EoS was analyzed in two steps. First, we computed the static EoS parameters using several XC functionals within the Rutgers-Chalmers scheme but different choices of exchange energy. Second, the VDoS was calculated at several pressures to compute the Helmholtz free energy. Table \ref{tab1} shows the static and 300 K EoS parameters calculated using these functionals. The usual performance of the standard LDA and PBE functionals is evident, in which the former underestimates and the latter overestimates the static volume $V_0$ with respect to the 300 K experimental value.
\begin{table*}[!t]
 \begin{center}
\caption{Equation of state (EoS) parameters computed in this study with LDA, PBE, and the vdW functionals: revPBE, optPBE, optB88, and optB86b. Quantum Espresso (QE), VASP, along with Ultrasoft (US) Pseudopotentials and Projector-Augmented Wave (PAW) datasets, and the all electron (AE) Wien2k software were used. The quasiharmonic approximation (QHA) was used to compute free energies and EoS parameters at 300 K, i.e., equilibrium volume, $V_0$, (\AA$^3$), bulk modulus, $K_0$ (GPa), bulk moduli derivative, $K'_0$ (dimensionless), and differences with respect to experiment, $\sigma V_0$ (\%). Experimental data at 300 K are from reference \cite{Boehler1980a}.}
\begin{tabular}{c c c c c c c}
 \hline\hline
 Code  & Functional    & $V_0$ & $K_0$ & $K_0'$ & Type   & $\sigma V_0$   \\
 \hline
QE     & PBE (static)          & 46.28 & 24.0  & 4.52   & US               &   3.1   \\
QE     & LDA (static)          & 40.78 & 32.2  & 4.65   & US               &   9.1   \\
\hline
Wien2k & PBE (static)          & 46.30 & 24.3  & 4.73   & AE               &   3.2   \\
Wien2k & LDA (static)         & 41.03 & 32.8  & 4.78   & AE               &   8.5   \\
\hline
VASP   & PBE (static)          & ~~45.22~~ & ~~24.3~~   & ~~4.64~~ & PAW              &   0.8   \\
VASP   & LDA (static)          & 40.09 & 33.3  & 4.79   & PAW              &  10.6   \\
\hline
VASP   & revPBE (static)       & 46.36 & 24.5  & 4.53   & PAW              &   5.0 \\
VASP   & optPBE (static)       & 44.50 & 26.8  & 4.59   & PAW              &   0.8 \\
VASP   & optB88 (static)       & 43.30 & 28.4  & 4.66   & PAW              &   1.6 \\
VASP   & optB86b (static)      & 43.33 & 27.6  & 4.67   & PAW              &   1.3 \\
\hline
VASP   & PBE (300 K)   & 47.54 & 19.1  & 4.86   & PAW + QHA        &   6.6   \\
VASP   & LDA (300 K)   & 41.85 & 25.5  & 5.26   & PAW + QHA        &   6.0   \\
VASP   & optB88 (300 K)& 44.99 & 23.5  & 4.85   & ~~PAW + QHA~~    &   0.3 \\
VASP   & optB86b (300 K)&45.01 & 23.2  & 4.82   & PAW + QHA        & 0.4 \\
\hline
VASP   & ~~optB88 + revPBE (300 K)~~& 44.84   & 24.2  & 4.77 & PAW + QHA        &   ~~0.02~~ \\
\hline \hline
Expt.  & (300 K)           & 44.85 & 25.8  & 4.37 &        $-$       &    -  \\
  \hline\hline
 \end{tabular}
\label{tab1}
\end{center}
\end{table*}
Except for revPBE results, all vdW functionals improve the static volume compared to LDA and PBE. A proper comparison with experiments should take into account thermal effects; therefore we assessed the performance of these functionals for vibrational frequencies by comparing the theoretical phonon dispersions with experimental ones at ambient conditions. Figure \ref{fig1} shows that LDA and PBE underestimate the frequencies and vdW interaction improves the agreement with experiments. In particular, the revPBE functional seems to produce the best agreement with experiments. 

Subsequently, the VDoS was computed at each volume using the same functional adopted for each static calculation, and the Helmholtz free energy was obtained. This procedure contrasts with that in reference \cite{Marcondes2015a}, where the PBE VDoS was used in all cases. However, the goal of that calculation was to develop a scheme to combine \textit{ab~initio} results with experimental compression curves to produce thermal EoSs with optimal accuracy to be used as pressure scale. According to table \ref{tab1}, after including vibrational effects, the LDA and PBE functionals produce similar errors of about 6\%, though in opposite directions. In contrast, the 300 K equilibrium volumes computed with the vdW functionals have an impressive agreement with experiments. For instance, the vdW optB88 functional reduces the error in $V_0$ to 0.3\%. Improvements in lattice parameters and bulk moduli of solids have previously been reported in static calculations using optB88 and optB86b functionals \cite{Klimes2011}. However, as shown here, this agreement further improved after inclusion of vibrational effects. 

\begin{figure}
\begin{center}
\includegraphics[scale=0.50]{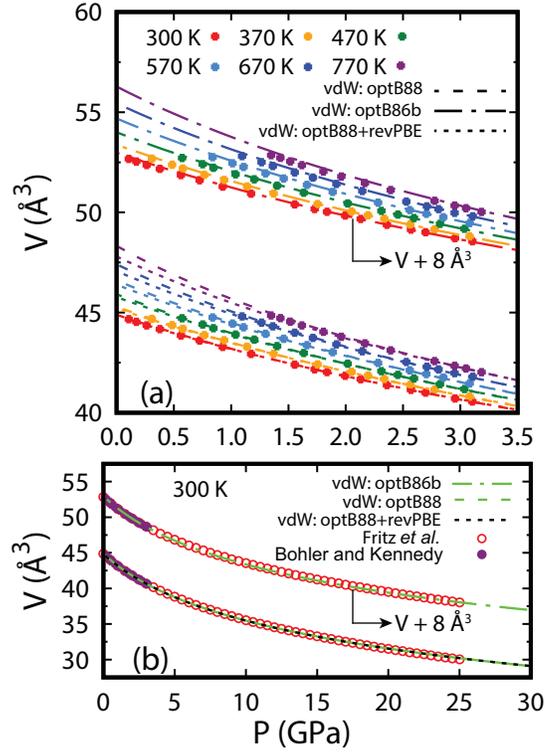}
\caption{\label{fig2} (a) High temperature compression curves of $B1$-type NaCl computed with optB88, optB86b, and the optB88+revPBE vdW functionals, compared to experimental data  from Bohler and Kennedy \cite{Boehler1980a}; (b) 300 K compression curve computed with the optB88, optB86b, and optB88+revPBE vdW functionals compared with the Hugoniot fitting from Fritz \textit{et al.} \cite{Fritz1972a}, and data from Bohler and Kennedy \cite{Boehler1980a}. The optB86b results are shifted by 8 \AA$^3$ for better visualization.}
\end{center}
\end{figure}

As seen in table 1, EoS parameters obtained using the optB88 and optB86b vdW functionals agree best with experimental data at 300 K, while phonons calculated using the revPBE functional agree best with experimental data, as shown in figure  \ref{fig1}. While accessing the performance of different functionals it is important to maintain consistency using the same functional for the entire calculation, practical tests often use a single VDoS combined with different static calculations. If this is the case, there is advantage in testing the performance of different functionals for phonon calculations and selecting the best performer for high temperature tests. Here, we combined the revPBE with the optB88 functional, but similar results are expected if the optB86b is used. Table 1 shows EoS parameters obtained after combining optB88 static results with the revPBE VDoS (optB88+revPBE). This procedure improves even further the overall agreement with experiments.  As seen in figure \ref{fig2}, compression curves of NaCl at high temperatures obtained using the optB88 functional for static calculations and revPBE for VDoS agree better with high temperature data than strictly optB88 of optB86b calculations.
As seen in figure \ref{fig2}(a), the optB88, optB86b, and optB88+revPBE compression curves at 300 K show excellent agreement with experiments. 
However, optB88 and optB86b results start deviating with increasing temperature, particularly at low pressures. If the agreement at 300 K is good, as it is here, this deviation is usually attributed to anharmonic effects, i.e., phonon-phonon interaction. A poor description of the XC energy usually manifests as errors at low temperatures also. As shown multiple times \cite{Marcondes2015a,Wu2008}, the difference between experimental data and calculated results at low temperatures can be removed and improve results at high temperatures, not only for the compression curve but for all thermodynamics properties \cite{Wu2008,Marcondes2015a}. Therefore, a clear manifestation of anharmonic effects is ambiguous because it depends on the performance of the XC functional being used. 

\begin{figure*}
\begin{center}
\includegraphics[scale=0.35]{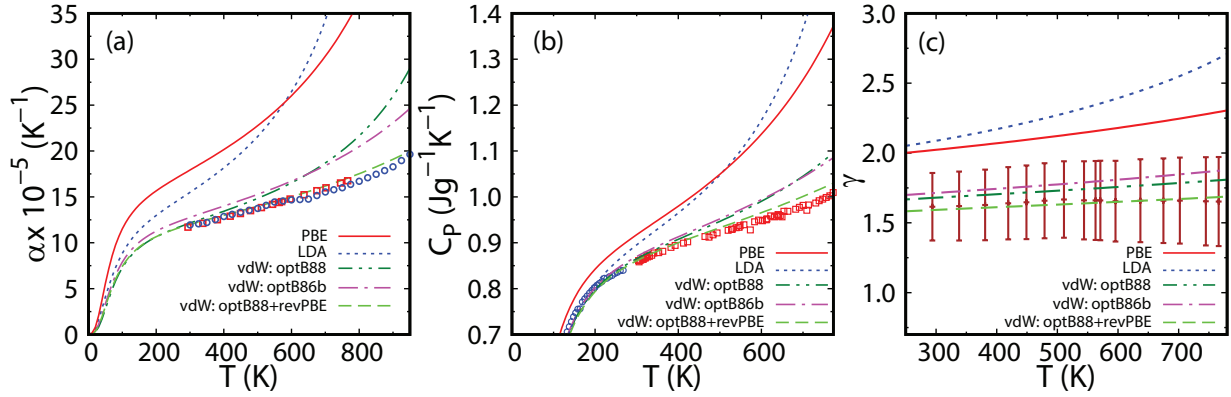}
\caption{\label{fig3} Thermodynamics properties of $B1$-type NaCl. (a) Thermal expansivity computed with the LDA, PBE, optB88 vdW, optB86b vdW and optB88+revPBE vdW functionals compared with data from references \cite{Anderson1987}  and \cite{Rao2013}; (b) Constant pressure specific heat computed with LDA, PBE, optB88 vdW, optB86b vdW,  and  optB88+revPBE vdW functionals, compared with experimental data from references \cite{Leadbetter1969} and \cite{Barron1964}; (c) Gr{\"u}neisen parameter computed with the LDA, PBE, optB88 vdW, optB86b vdW,  and  optB88+revPBE vdW functionals.  Experimental data are from reference \cite{Anderson1987}.}
\end{center}
\end{figure*}

\begin{figure}
\begin{center}
\includegraphics{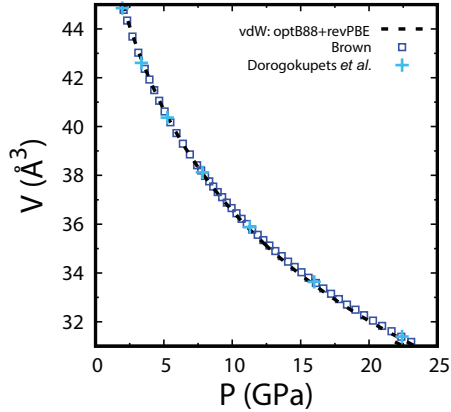}
\caption{\label{fig7} Compression curves of $B1$-type NaCl calculated with the optB88+revPBE vdW  functionals at 1000 K compared with semiempirical results by Brown \cite{Brown1999a} and Dorogokupets \textit{et al.} \cite{Doro2007}.}
\end{center}
\end{figure}

This effect is shown in figure \ref{fig3} comparing thermodynamics properties at high temperatures obtained with LDA, PBE, optB88, optB86b, and optB88+revPBE functionals with experimental data. First, it is clear that the vdW functionals improve greatly the agreement between calculated and measured thermodynamics properties, in all cases. Second, the thermal expansivity shown in figure \ref{fig3} (a) offers the most direct assessment of anharmonicity. At low temperatures, the thermal expansivity computed with the optB88 and optB86b functionals agree with experiments, but they start deviating from experimental data around 500 K. This is usually a tight estimate of the temperature at which QHA starts to break down \cite{Wu2008,Carrier2007,Wentzcovitch2004}. In contrast, the thermal expansivity computed with the optB88+revPBE shows an impressive agreement with experiments, up to the NaCl melting point at $\approx$ 1000 K at 0 GPa. Whether this should be the case is unclear. However, we point out here that the best functional for phonon calculations also produces the best results for high temperature thermodynamics properties. This has important implications for computations of anharmonic effects since they appear to be very sensitive to the choice of XC functional, particularly if vdW interaction is significant. 
Finally, figure \ref{fig7} shows the NaCl high temperature EoS computed with the optB88+revPBE  functionals compared to the semiempirical fitting from Brown \cite{Brown1999a}, Dorogokupets \textit{et al.}\cite{Doro2007}, and Matsui \textit{et al.} \cite{Matsui2012}, often used as pressure scales. The agreement is outstanding. While their results include experimental data, ours  were obtained solely using \textit{ab~initio} methods. 

\subsection{$B$1 - $B$2 structural phase transition}

\begin{figure}
\begin{center}
\includegraphics{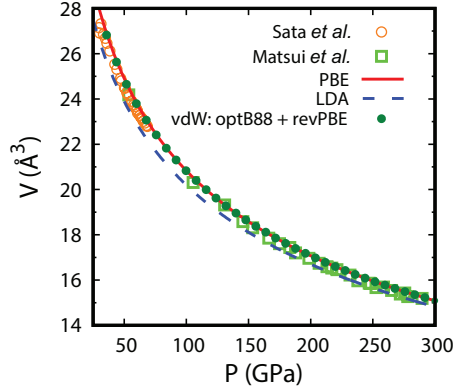}
\caption{\label{eosb2} 300 K compression curve of $B2$-type NaCl computed with the LDA, PBE, and optB88 + revPBE. Experimental data are from references \cite{Sakai2011} and \cite{Sata2002}.}
\end{center}
\end{figure}

NaCl exhibits a structural phase transition from the rock-salt ($B1$) to the CsCl structure ($B2$) \cite{Nishiyama2003,Li1987} close to 30 GPa. With increasing pressure, the error of the standard LDA and PBE functionals is reduced. Therefore, the $B2$-type NaCl EoS parameters at 300 K compare reasonably well with experimental values irrespective of the functional used, as shown in table \ref{tab2}. Although the LDA error seems large compared to those of other functionals, these results are extrapolated to 0 GPa after fitting the third order Birch-Murnaghan EoS. Understandably, vdW interaction does not affect the calculated $B2$ phase compression curve significantly. Indeed, PBE and the best combination of vdW functionals that describes the $B1$ phase, i.e., optB88+revPBE, produce similar results and in good agreement with experimental data (see figure \ref{eosb2}).

Static enthalpy difference between the $B1$ and $B2$ phases produces static transition pressures equal to 26, 29, and 28 GPa for LDA, GGA, and optB88+revPBE functionals, respectively. Direct comparison of the Gibbs free energies of these two phases decreases these transition pressures slightly, as expected for a transition with negative Clapeyron slope ($dP_{\rm tr}/dT$). Nevertheless, the agreement between calculated and measured phase boundaries is excellent and seemingly within the range of experimental uncertainties (see table \ref{tab2}).

\begin{figure}
\begin{center}
\includegraphics[scale=0.40]{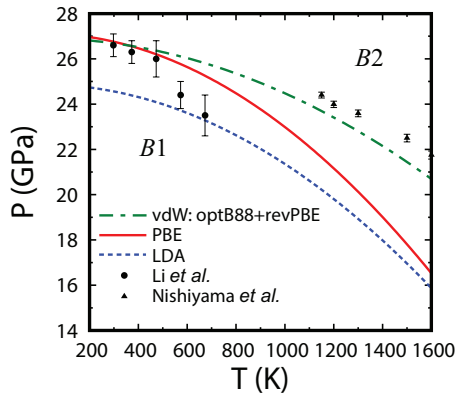}
\caption{\label{gibbs-b1-b2} $B1$ - $B2$ phase diagram of NaCl computed with LDA, PBE, and the optB88+ revPBE compared with experimental data from references \cite{Li1987, Nishiyama2003}.}
\end{center}
\end{figure}

\begin{table}[h!]
\begin{center}
\caption{EoS parameters ($V_0$ in \AA$^3$,  $K_0$ in GPa) of $B2$-type NaCl at 300 K. $B1$ - $B2$ phase transition  pressure ($P_{\rm tr}$) in GPa, and Clapeyron slope ($dP_{\rm tr}/dT$) in MPa/K, at several temperatures,
computed using LDA, PBE, and  optB88+revPBE vdW functionals.}
\label{tab2}
\begin{tabular}{|c|c|c|c|c|c|}
\hline\hline
\multicolumn{2}{|c|}{ }        & LDA      &    PBE     &   optB88+revPBE    &    Expt.  \\
\hline
\multicolumn{2}{|c|}{$V_0$ }               & 37.00    &    44.83   &        40.46       &     39.49$^{a}$ \\
\multicolumn{2}{|c|}{$K_0$ }               & 44.5     &    35.1    &        37.3        &     39.7$^{a}$  \\
\multicolumn{2}{|c|}{$K'_0$}               & 4.26     &    4.21    &        4.18        &     4.14$^{a}$  \\
\hline
\multirow{3}{*}{$P_{\rm tr}$} & 300 K               & 24.7  &   27.1    &   26.9    &    26.6$^{b}$  \\
                     & 600 K                        & 23.6  &   25.6    &   26.0    &    24.2$^{b}$  \\
                     &1200 K                        & 19.8  &   21.1    &   23.3    &    24.0$^{c}$  \\
\hline
\multirow{3}{*}{$\frac{dP_{\rm tr}}{dT}$} & 300 K  & -3.0   &   -4.5     &    -2.7    &    -3.4$^{b}$   \\
                     & 600 K                        & -4.7   &   -5.9     &    -3.5    &    -9.0$^{b}$   \\
                     &1200 K                        & -8.0   &   -8.8     &    -5.1    &    -5.2$^{c}$   \\
\hline\hline
 \end{tabular}
 \\
$^{a}$ Reference \cite{Sakai2011}, $^{b}$ Reference \cite{Li1987}, $^{c}$ Reference \cite{Nishiyama2003}
\end{center}
\end{table}
The phase diagram of NaCl computed using LDA, PBE, and optB88+revPBE functionals is shown in figure \ref{gibbs-b1-b2}. 
As anticipated, the best combination of vdW functionals also predicts the best phase boundary and Clapeyron slopes at all temperatures. Two points at 600 K in the phase boundary measured by Li \textit{et al.}\cite{Li1987} appear to be off since they deviate substantially from the high temperature data from Nishiyama \textit{et al.}\cite{Nishiyama2003}. The PBE phase boundary and Clapeyron slope agree well with experiments at lower temperatures but deviate rapidly from experiments at high temperatures. The Clapeyron slope at 300 K calculated with the LDA is consistent with that predicted by the optB88+revPBE calculation and with experimental data. However, LDA underestimates transition pressures and the entire phase boundary is shifted to lower pressures compared with the PBE and optB88+revPBE boundaries. This behavior has been previously reported numerous times \cite{Tsuchiya2004,Yu2008,Wentzcovitch2010a} and is expected. 

\section{Summary and Conclusions}
\label{sec4}

In this paper, we used several XC functionals to test the importance of the vdW interaction to describe structural, vibrational, and thermodynamics properties of NaCl. Compared to the standard LDA and PBE functionals, vdW interaction improves the description of these properties tremendously. High temperature properties, especially thermal expansivity, the property most sensitive to anharmonic effects were obtained with high accuracy using the QHA and the best performing combination of vdW functionals (optB88+RevPBE), even at temperatures near melting where anharmonic effects are expected to be significant. This surprising result indicates that anharmonic effects, usually manifested in the deviation between measurements and QHA results for the thermal expansivity are clearly quite sensitive to the choice of XC functional used, especially if vdW interaction is important. Low temperature errors in the compression curve caused by DFT can be perceived as anharmonic effects at high temperatures owing to the very construct of the QHA. Therefore, improving the compression curve at low temperatures and phonon frequencies’ dependence on volume should produce more accurate thermodynamic properties and a better description of intrinsic anharmonic effects. 

The best performing combination of vdW functionals, optB88+RevPBE, also reproduces the $B1$ to $B2$ phase boundary at high temperatures very well, something that cannot be accomplished with LDA or PBE functionals. Present results clearly indicate that vdW interaction is key to improving the description of the ionic bond in NaCl and likely in other alkali halides. Detailed calculations in other similarly ionic materials are needed to further validate the generality of our conclusions.

\vspace{0.2cm}

\textbf{ACKNOWLEDGMENTS}

MLM and LVCA thank support by the National Council for Scientific and Technological Development (CNPq). MLM also thanks support by Coordena\c{c}\~ao de Aperfei\c{c}oamento de Pessoal de N\'ivel Superior (CAPES - grant N$^\circ$ BEX 14456/13-3) via the science without borders program. RMW and MLM were supported by NSF grants 1341862 and 1348066. Calculations were performed in the Texas Advanced Computing Center (Stampede and Stampede2) under an XSEDE allocation, and in the Blue Waters at UIUC.

\section*{References}
\bibliographystyle{nature}

\end{document}